\documentclass{article}

\usepackage{graphicx}
\usepackage{amsmath, amssymb}
\usepackage{natbib}
\usepackage{booktabs}

\begin{document}

\title{Modeling Credit Spreads Using Nonlinear Regression}
\author{Radoslava Mirkov \and Thomas Maul \and Ronald Hochreiter \and Holger Thomae}
\date{July 2013}

\maketitle

\begin{abstract}
\noindent The term structure of credit spreads is studied with an aim to predict its future movements. A completely new approach to tackle this problem is presented, which utilizes nonlinear parametric models. The Brain-Cousens regression model with five parameters is chosen to describe the term structure of credit spreads. Further, we investigate the dependence of the parameter changes over time and the determinants of credit spreads.
\end{abstract}


\section{Introduction and Motivation}

We study the historical development of the credit spread curves, and are interested in forecasting future movements of credit spreads. In economic sciences, credit spreads represent the premium paid for specific risks embedded in a bond. The risk factors include geopolitical and macroeconomic variables. 
For details, see e.g. \citet{Schlecker2009}.
The existing methods used in the banking industry proved unsatisfactory in times of financial crisis, as the relationship between issuer and reference curves has changed. 

We scrutinize the behavior of credit spreads from a completely different perspective. 
We model  the credit spread curve not by the common layer-factor approach, but we approximate the curve by a nonlinear parametric function with several parameters.  Then we concentrate on finding the dependencies between these parameters and timely available and observable indicators and market data. This is motivated by the fact that the complexity of the parametric curve can be reduced to a small number of parameters so that changing patterns of the curve structure can be understood in terms of changes in these parameters. Also, each nonlinear parametric curve may be summarized by its parameter estimates as a single low-dimensional multivariate observation, which then may be subject to a regression or a correlation analysis.

The following model which describes the structure of credit spreads $y_i$ for given times to maturity $x_j$ is studied:
$$ y_i =  y(x_j) + \varepsilon_i, $$
where $x_j, \, j = 1, \ldots, 12$, denotes the time to maturity in years of quoted credit spreads, usually 
$ x = (1, 2, 3, 4, 5, 6, 7, 8, 9, 10, 15, 20) ,$ 
and $\varepsilon_i \sim \mathcal F (0, \sigma^2)$ are error terms with zero mean and constant variance $\sigma^2$, for  $i = 1, \ldots, n$.

\section{Data Description}

\begin{figure}[bt!]\centering
\includegraphics[width=5.5cm]{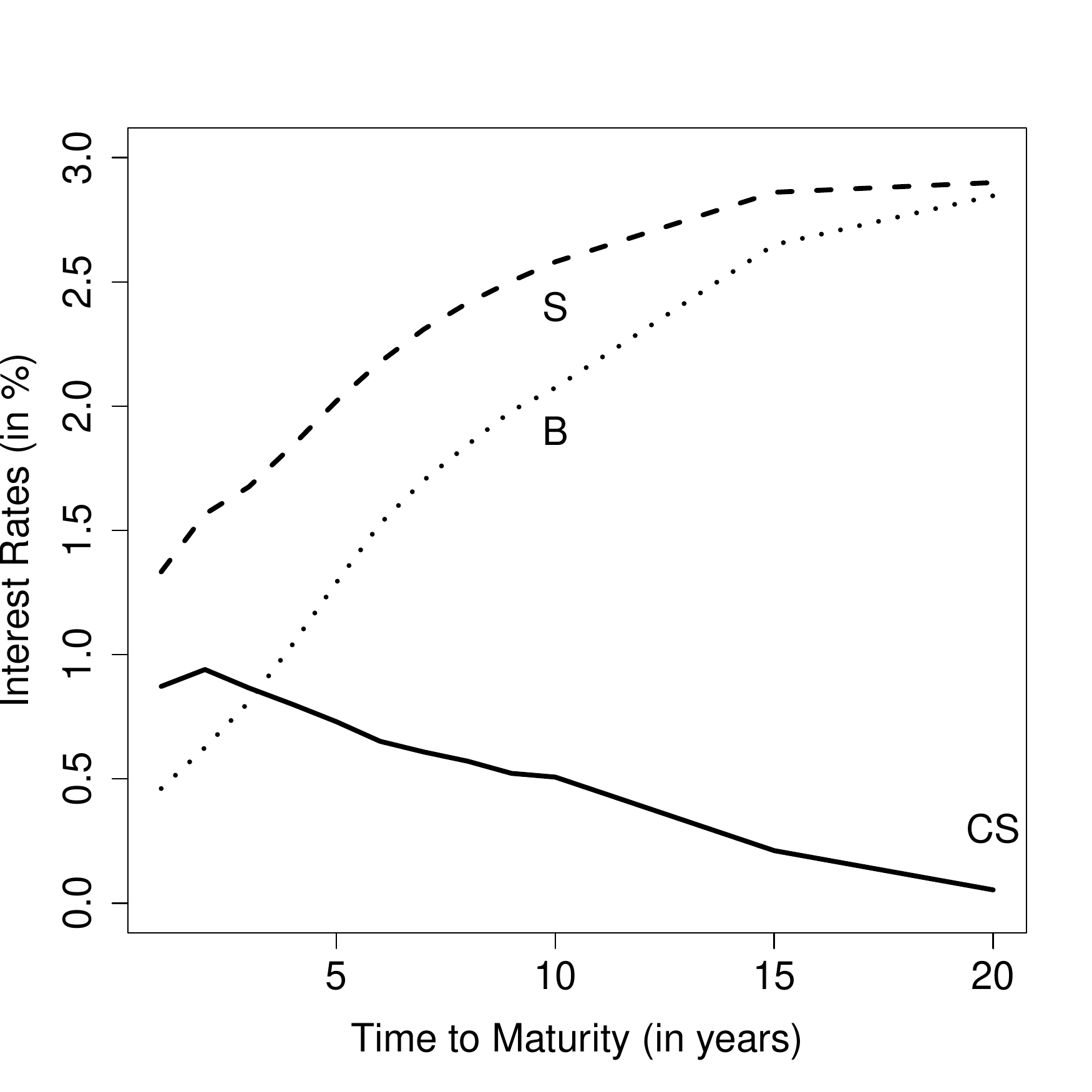} \hspace{0.1cm}
\includegraphics[width=5.5cm]{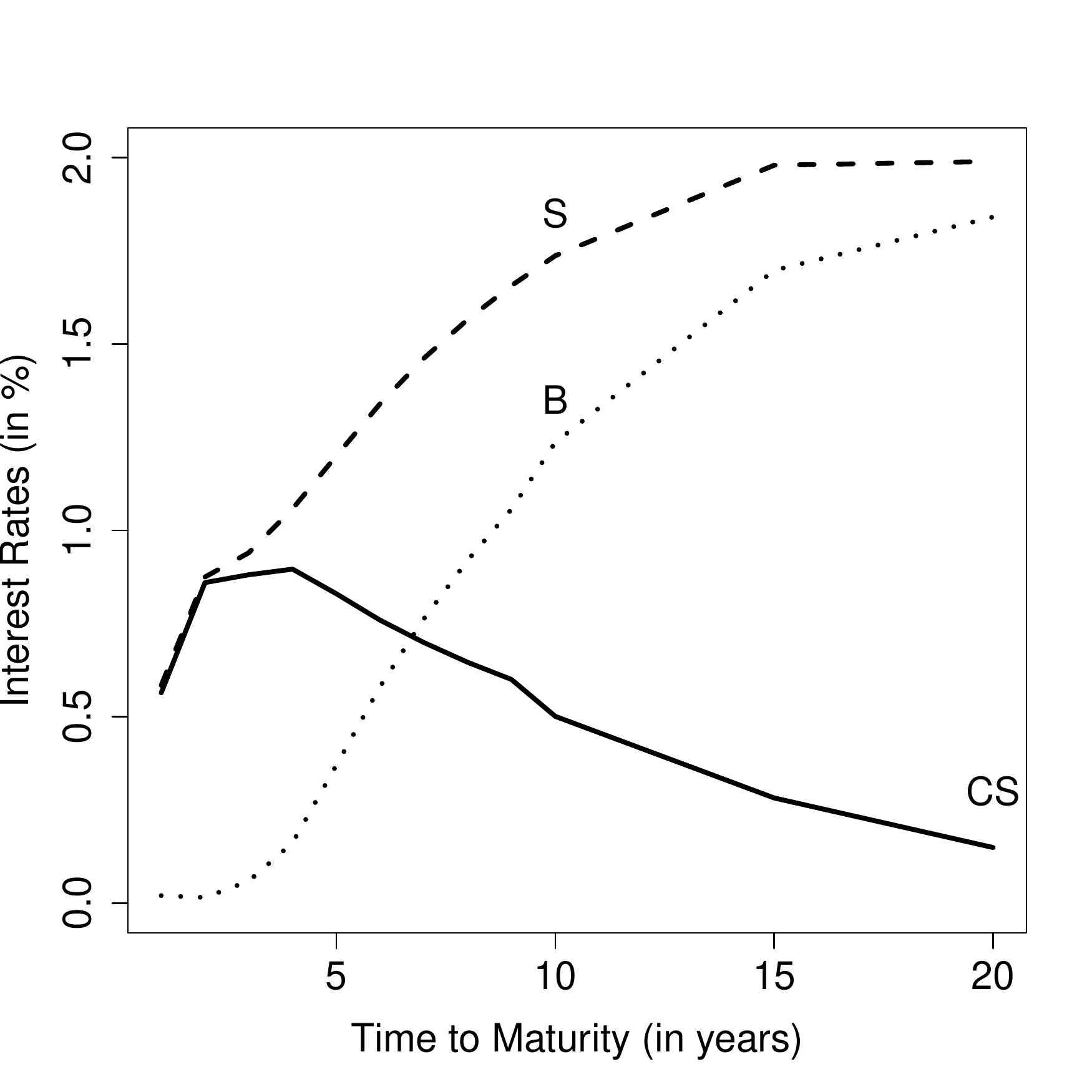}
\caption{\label{mirkov:fig1} The term structure of ISDA fixings ($S$, dashed line), German government bonds ($B$, dotted line) and credit spreads ($CS$, solid line) on October 21, 2011 (left) and May 31, 2012 (right).}
\end{figure}

The analyzed data set is based on an excerpt from the Bloomberg data base, and contains daily quotation of ISDA fixings for Euro and German government bonds for all maturities for the period between June 2011 and June 2012, i.e. $n=258$. 
The credit spreads $y_i$ are obtained by subtracting the issuer curve from the reference curve, as Figure~\ref{mirkov:fig1} shows graphically. Comparing credit spread curves in Figure~\ref{mirkov:fig1} (left) and (right), it becomes evident that the structure of credit spread curves is changing, and we need a model that will be flexible enough to capture possible developments and different shapes of credit spread curves. 

\section{Nonlinear Regression Model}

We fit a parametric nonlinear logistic regression model and analyze the term structure of credit spreads. The generalization of the log-logistic regression, the so-called Brain-Cousens model (BC-model) is proposed by \citet{Ritz2008} for this kind of dependency. Some authors propose some variant of spline regression for similar problems, see e.g. \citet{Jarrow2004}.

The BC-model is defined by
\begin{equation} \label{mirkov:eq1}
   y (x_j) = c + \frac{d + f \, x_j - c}
		{1 + \exp ( b \, [\ln (x_j) - \ln (e)] )},
\end{equation}
and the parameters in model~(\ref{mirkov:eq1}) have the following meaning: $c$ and $d$  define the upper and lower horizontal asymptotes,  $f$ is the slope of the upper asymptote, while $b$ and $e$ describe the shape of the decrease of the curve, i.e. $e$ is the inflection point of the curve, and $b$ is proportional to the slope at $x_j = e$.  A similar approach for describing nonlinear dependencies is proposed in \citet{Friedl2012}. 

The starting values for the iteration necessary to calculate the least squares estimates of parameters are obtained either by using the parameter estimates of the previous day or by generating random starting values. 
If the parameter estimates from the previous day used as the starting values for the next day's estimation did not lead to convergence, random starting values from the interval $[-2,2]$ are used. This approach is introduced in order to obtain convergence also in cases when the behavior of credit spreads changes dramatically from one day to another. It also enables the identification of days when market fluctuations influence the development of credit spreads strongly. 

\begin{figure}[bt!]\centering
\includegraphics[width=5.5cm]{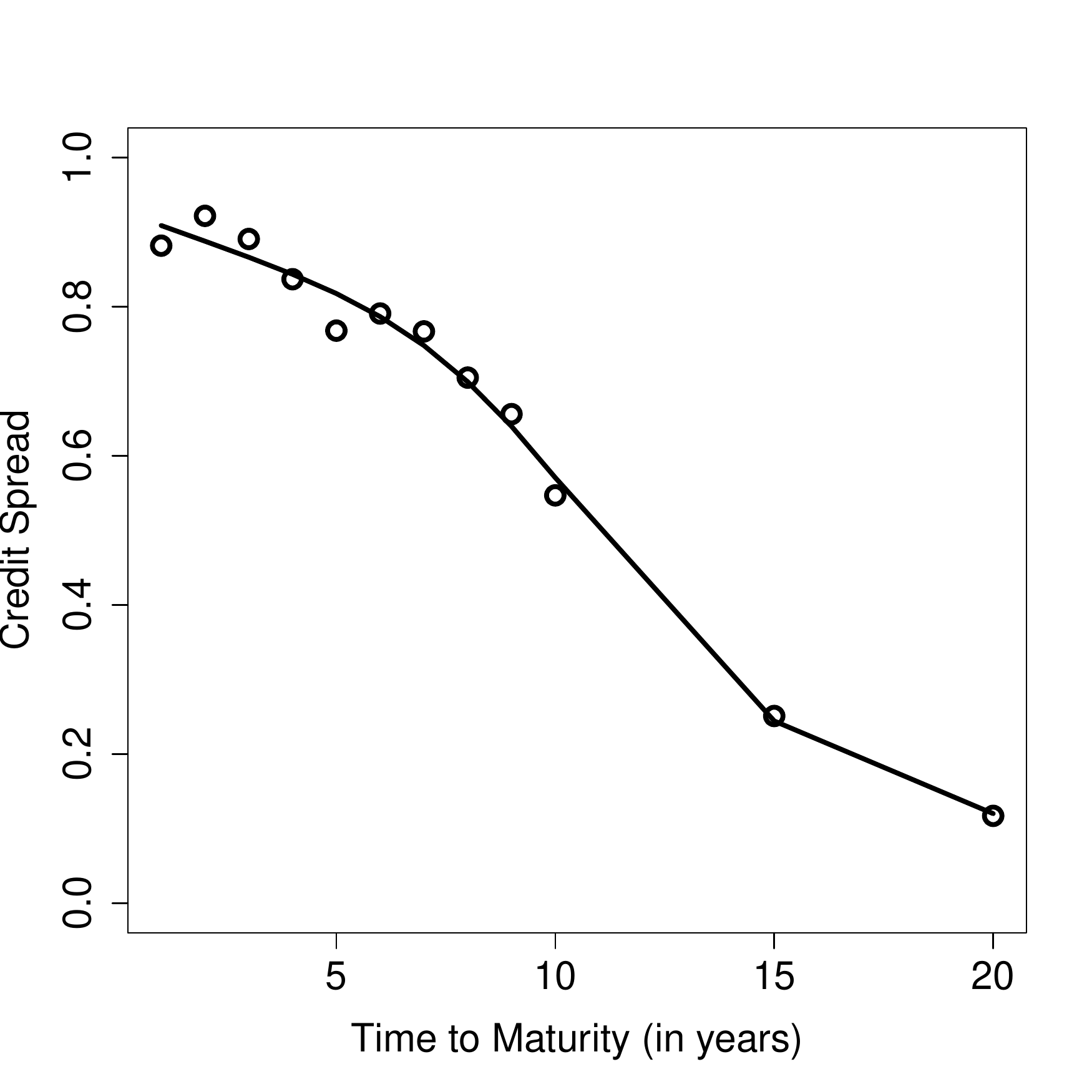} \hspace{0.1cm}
\includegraphics[width=5.5cm]{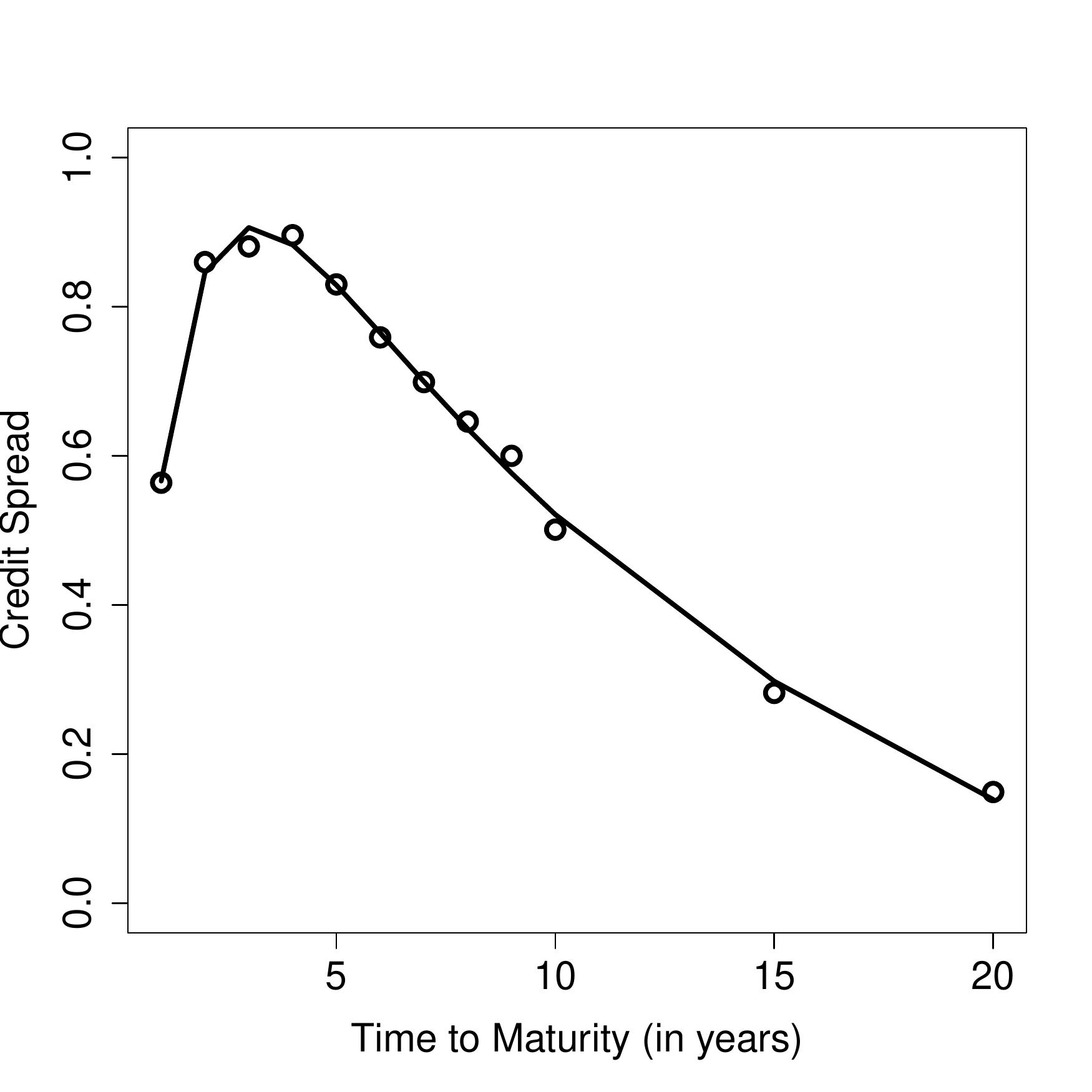}
\caption{Fitted BC model given by (\ref{mirkov:eq1}) (solid line) and quoted credit spread values at time to maturity on October 21, 2011 (left) and May 31, 2012 (right).}
\label{mirkov:fig2}
\end{figure}

The results of the parameter estimation for both curves shown in Figure~\ref{mirkov:fig1} are given in Table~\ref{mirkov:tab}. We refer to Figure~\ref{mirkov:fig2} for a graphical representation of the fitted models. We note that for the model fit shown in Figure~\ref{mirkov:fig2} (left), nine random iterations of starting values were necessary to obtain  convergence. This method yields convergence for 231 out of 258 days. Without random starting values, the convergence is obtained for only 180 days.

\begin{table}[!hb] \centering
\caption{\label{mirkov:tab} MLEs (std. errors) of the BC-model (left) and (right).}
\medskip
\begin{tabular}{ccccc}
\toprule[0.09 em]
b & c & d & e & f \\
\midrule
4.9129   &  0.0772  & 0.9300 & 12.7469 & -0.0209 \\
(1.3833)  & (0.0648)  & (0.0307) & (1.0744) & (0.0094) \\
\midrule
1.3550   &  -1.5464  & -0.2754 & 2.4529 & 1.4679 \\
(0.1951)  & (1.1451)  & (0.3344) & (0.8448) & (1.0474) \\
\bottomrule[0.09 em]
\end{tabular}
\end{table}

\section{Conclusion}

We study the term structure of credit spreads with an aim to predict their future movements. 
We suggest a completely new approach to tackle this problem, and instead of modeling credit spreads by the means of the layer-factor model, we utilize a nonlinear parametric model and concentrate on its parameters. The Brain-Cousens regression model with five parameters is chosen to describe the term structure of credit spreads.
Random starting values are introduced in order to obtain convergence of parameter estimates also in cases when the behavior of credit spreads changes dramatically. Eventually, the dependence of the parameter values and given microeconomic factors over time is to be analyzed. 

\bibliographystyle{plainnat}
\bibliography{iwsm13}

\end{document}